\documentclass[10pt,twocolumn,twoside,letterpaper]{IEEEtran}

\usepackage{cite}
\usepackage{amsmath,amssymb,amsfonts}
\usepackage{graphicx}
\usepackage{siunitx}

\usepackage{lineno}

\usepackage{geometry}
\makeatletter
\def\ps@IEEEtitlepagestyle{
  \def\@oddfoot{\mycopyrightnotice}
  \def\@evenfoot{}
}
\def\mycopyrightnotice{
  {\footnotesize
  \begin{minipage}{\textwidth}
  \centering
  978-1-6654-2113-3/21/\$31.00 \copyright2021 IEEE
  \end{minipage}
  }
}

\usepackage{url}

\begin{document}
\title{Commissioning and installation of the new small-Diameter Muon Drift Tube (sMDT) detectors for the Phase-I upgrade of the ATLAS Muon Spectrometer
}
%
%
%

\author{G. H. Eberwein, O. Kortner, H. Kroha, M. Rendel, P. Rieck, D. Soyk, \\ E. Voevodina$^1$ and V. Walbrecht on behalf of the ATLAS Muon Group 

\thanks{Manuscript received December 1, 2021.}
\thanks{$^1$ Corresponding author. (e-mail: voevodin@mpp.mpg.de).}
\thanks{G. H. Eberwein was with Max-Planck-Institute for Physics, Foehringer Ring
6, 80805 Munich, Germany. He is now with Department of Physics, University of Oxford, Denys Wilkinson Building, Keble Road, Oxford OX1 3RH, UK.}
\thanks{O. Kortner, H. Kroha, E. Voevodina, M. Rendel, P. Rieck, D. Soyk and V. Walbrecht are with Max-Planck-Institute for Physics, Foehringer Ring 6, 80805 Munich, Germany.}}

\maketitle

\pagenumbering{gobble}

\begin{abstract}
The Monitored Drift Tubes, as a part of the ATLAS muon spectrometer, are precision drift chambers designed to provide excellent spatial resolution and high tracking efficiency independent of the track angle. Through the life of the LHC and ATLAS experiment, this detector has already demonstrated that they provide precise tracking over large areas. The aim of the ATLAS muon spectrometer upgrade is to increase the muon trigger efficiency, precise muon momentum measurement and to improve the rate capability of the muon system in the high-background regions during the High-Luminosity LHC runs. To meet these requirements, the proposed solution is based on the small (15 mm) diameter Muon Drift Tube chamber (sMDT) technology. The new detector provides about an order of magnitude higher rate capability and allows for the installation of additional new triplet Resistive Plate Chambers (RPCs) trigger detectors in the barrel inner layer of the muon system. A pilot project for the barrel inner layer upgrade is underway during the 2019/21 LHC shutdown. For this reason, the Max-Planck-Institut f{\"u}r Physik in Munich has built 16 sMDT chambers, each will cover an area of about 2.5 $m^2$. To ensure their proper operation in the experiment, the sMDT detectors have to pass a set of stringent tests both at the production site and after their delivery at CERN. After their installation in the ATLAS muon spectrometer, the muon stations are further tested and commissioned with cosmic rays. The author will describe the detector design, the quality assurance and certification path, as well as will present the experience with the chamber tests, the integration procedure and installation of the muon stations in the ATLAS experiment.
\end{abstract}


\section{Introduction}
\begin{figure}
\centering
\includegraphics[width=8.5cm]{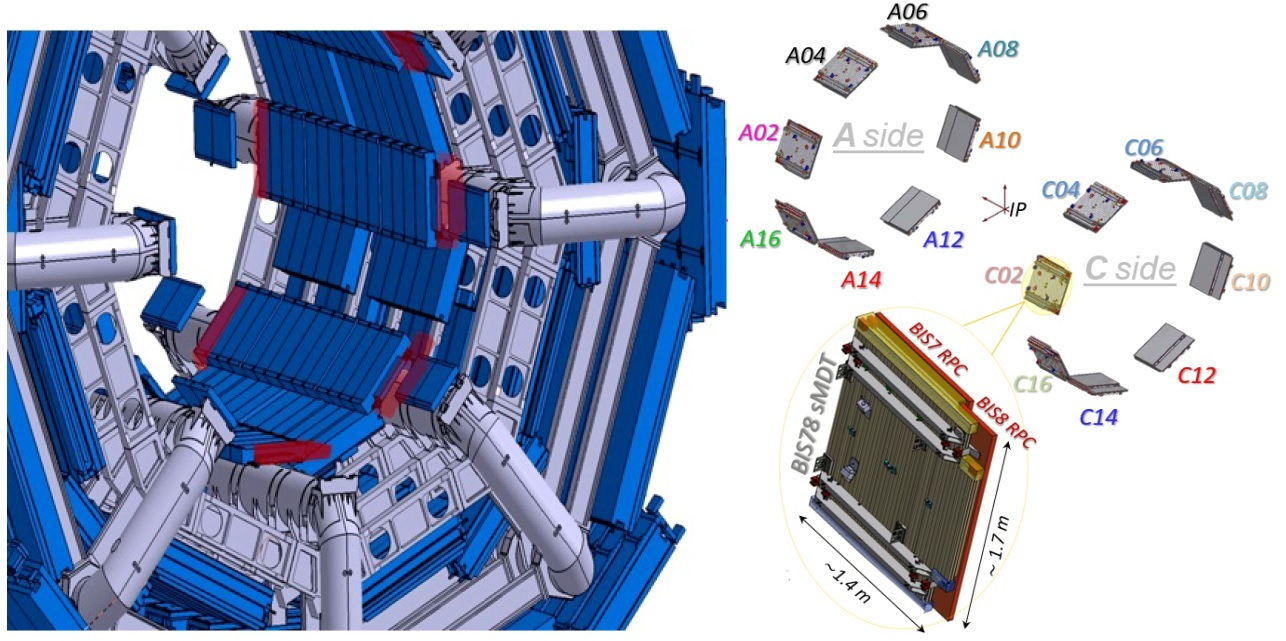}
\caption{Schematic of the BIS78 project for the Phase-1 upgrade of the ATLAS muon spectrometer.}
\label{fig1}
\end{figure}

%
%
%
%
\IEEEPARstart{T}{he} success of the muon spectrometer is based on the high muon trigger efficiency and muon momentum resolution up to the TeV scale. However, in the innermost transition region between the ATLAS barrel and the endcap ($1.0<|\eta|<1.3$), a non-negligible rate of the fake muon trigger has been observed due to the high rate based on secondary charged tracks generated by beam halo protons and a lack of detector instrumentation.  To reinforce the fake rejection and to improve the selectivity of the muon trigger in this problematic region, the two new generations of the gaseous detector technology, such as the triplet thin gap Resistive Plate Chambers (RPC) and small (15 mm) diameter Muon Drift Tube (sMDT) detectors, will be installed during the LHC Long Shutdown 2 (LS2, 2020-2021). That Phase-I upgrade serves as a pilot project, known as BIS78 (shown in Fig.~\ref{fig1}), for the complete replacement of the existing 96 BIS1-6 MDT chambers in the LS3 period (2025-2027) by the combined sMDT+RPC muon modules that will have a very similar design\cite{b1}. The first 16 new BIS78 sMDT detectors are currently under installation and commissioning at CERN: this smaller size project mainly aims to validate the final mechanical installation procedure, the correctness of the services previously installed in the experimental cavern, and to obtain early feedback about the front-end electronics response and detection performance in the ATLAS environment with the detectors powered and read out through the final services.

\section{sMDT Detector Design}
Each individual BIS78 muon station serves as the integrated tracking and triggering detector and therefore composes of one sMDT chamber with two different sizes of RPC triplets (BIS7 and BIS8). Furthermore, the sMDT detectors consist of two multilayers of the high pressurized drift tubes, each containing four layers of densely packed tubes mounted on an aluminum support frame: the chamber sizes vary from about 2.67 $m^2$ to 3.04 $m^2$ while tube lengths range from 1009 mm to 1669 mm. In order to uniformly fill the space by taking into account the features of the toroid magnet coil at the ends of the barrel inner layer and to reduce the number of dead areas, 12 different sMDT detectors versions have been properly designed in MPI Munich and approved by ATLAS Muon Community. For detector control data, the BIS78 sMDT chambers are equipped with 12 temperature sensors distributed on the multilayer and the support structure. On the top surface of the detector are typically mounted eight platforms: four Axial-Praxial, two CCC and two B-field platforms. In addition, each detector is instrumented with the in-plane alignment system, consisting of two image sensors, four lenses, and two or four masks, in order to monitor the deformations such as sag and torsion. For the 16 A/C-side BIS78 sMDT detectors, the chamber-to-chamber alignment system and B-field components will be re-used from the previous MDT detectors\cite{b2}.

\section{Validation Tests and Integration at CERN}
\begin{figure}[htbp]
\centerline{\includegraphics[width=8.5cm]{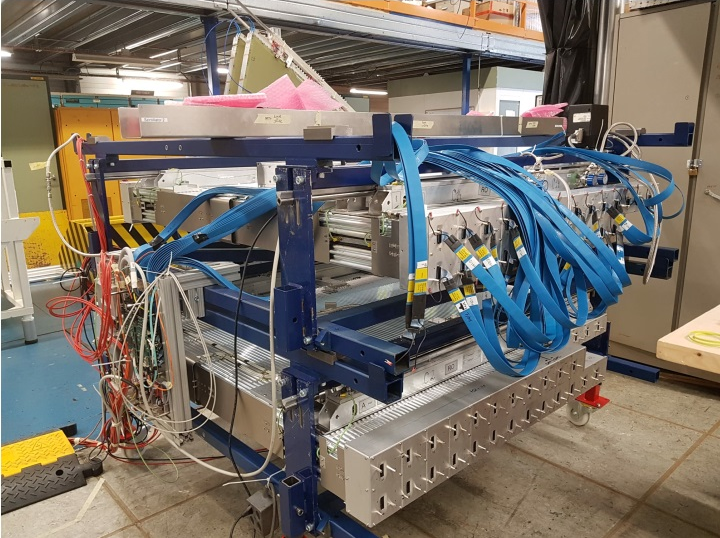}}
\caption{Picture of the new sMDT chamber during the commissioning campaign at CERN BB5 facility.}
\label{fig2}
\end{figure}

Upon sMDT chambers arrival at CERN BB5 construction facility, each detector is subjected to a strict validation test program: gas tightness, connectivity, noise rate measurement of the front-end electronics and test with cosmic rays, in order to detect any damages which might have occurred during the uncontrolled conditions of the transport from the MPI Munich production site and to ensure their proper operation in the ATLAS muon spectrometer. A picture of the sMDT detector during the commissioning test campaign at CERN BB5 facility is shown in Fig.~\ref{fig2}.

The leak rate of each multilayer is determined from the temperature corrected pressure drop over a time period of at least 24 hr. In order to avoid the possible gas contamination in the chamber, since it significantly can change the maximum drift time, detectors are evacuated and then filled with the operating $Ar/CO_{2}$ (93:7) gas mixture at the pressure 3 bar. The maximum acceptable gas leak rate is $2N \times 10^{-8}$ bar$\times$l/s for N tube in a detector. The measurement accuracy achievable after 24 hr obtained during the commissioning campaign is 0.063 mbar/h. The high voltage stability at the nominal operating high voltage (HV) point 2.73 kV is tested by measuring the dark current of each tube layer. A noise rate test is performed without and with HV to easily distinguish between electronic noise and tubes with internal discharges. The average intrinsic noise rate of all tested chambers has been found of $(0.10 \pm 0.02)$ kHz/tube at the operating threshold value of -39 mV and at the working point of 2.73 kV, well below the upper noise limit of the sMDT detectors installed in the ATLAS muon system, that is above 5 kHz/tube. Moreover, the control interface of the front-end electronics and communication stability are checked as well. A complete system test of the sMDT detector is performed with cosmic muons. The maximum drift time spectra, spatial resolution and muon detection efficiency for each tube of the sMDT chamber are computed and recorded as well as the detector hit profile is measured to allow a check of the correct channel assignment.

\begin{figure}[htbp]
\centerline{\includegraphics[width=8.5cm]{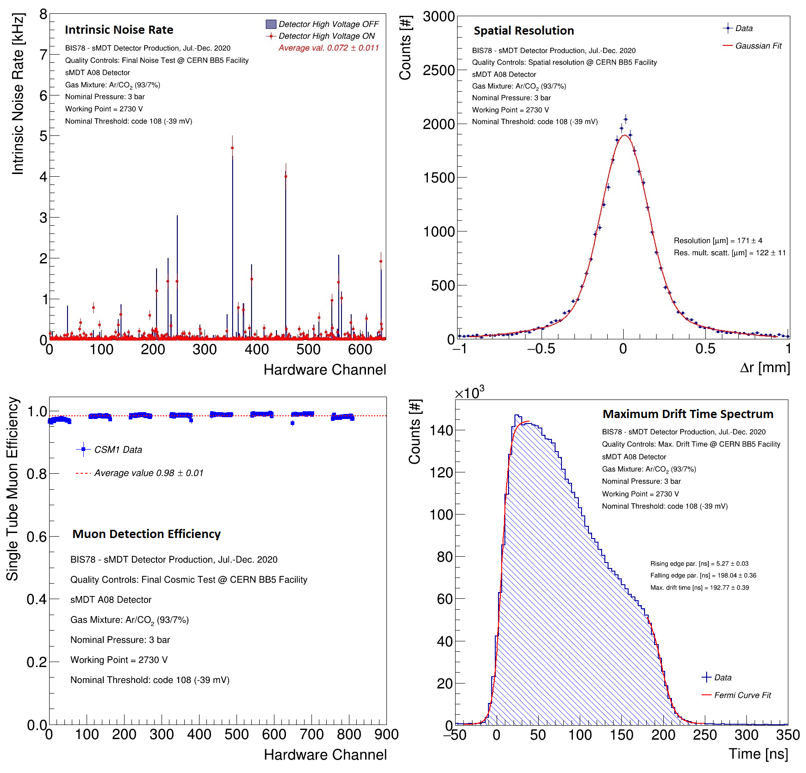}}
\caption{BIS78 sMDT performance study during the validation tests at CERN BB5 facility.}
\label{fig3}
\end{figure}

The  MDT  chamber under the commissioning test reaches an efficiency of 98.1$\%$ for muons when operated with $Ar/CO_{2}$ (93:7) gas mixture, HV = 2.73 kV and has a read-out with all mezzanines,  configured with -39 mV threshold.  The average chamber efficiency is in an agreement both with the previous QA/QC test campaign a the MPI production site and with the Monte Carlo (MC) simulation prediction.  The spatial resolution for the BIS78 sMDT detector under validation test is measured to be (122 $\pm$ 11) $\mu$m with the multiple scattering correction obtained from Monte Carlo simulation, $\sigma_{MC}$ = (120 $\pm$ 10) $\mu$m \cite{b3}.  This performance is fully in agreement with the previous measurement at the muon beam facility, $\sigma_{\mu}$ = (125 $\pm$ 4) $\mu$m \cite{b4}. Examples of typical results can be seen in Fig ~\ref{fig3}.

\begin{figure}[htbp]
\centerline{\includegraphics[width=8.5cm]{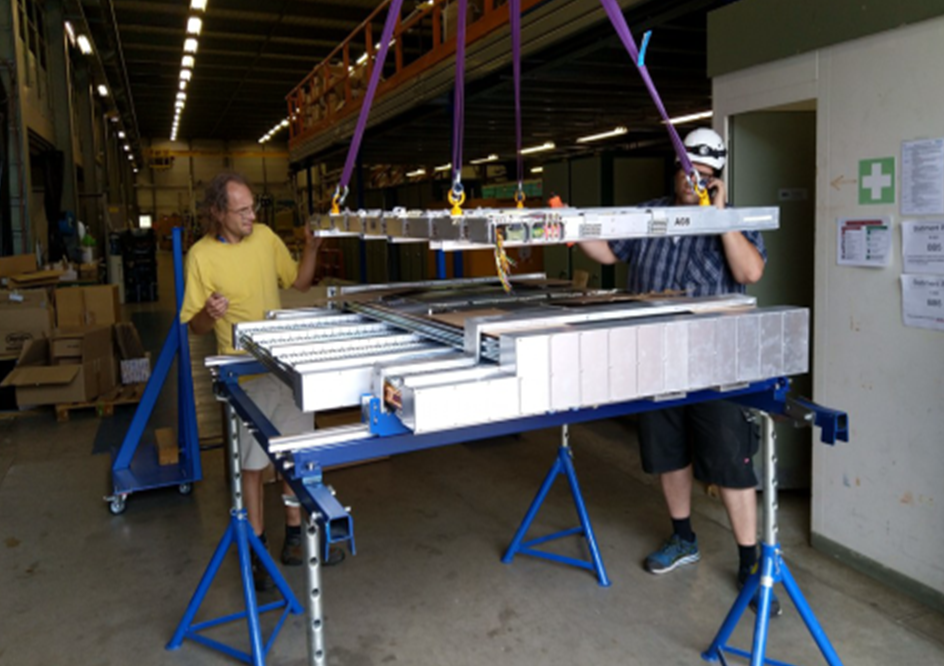}}
\caption{Photograph of the sMDT and RPC detectors integration process.}
\label{fig4}
\end{figure}

After the commissioning campaign, the sMDT chamber is then integrated with its two triplet RPC trigger detectors to form a so-called final muon station. The challenges for the mechanical design of the BIS sMDT are the small available space that needs to share with the new RPC detectors and minimizing the gravitational deformation of the thin chambers with minimal space available for the support structures. For this reason, the BIS78 muon stations require a special common support frame to carry the RPC and sMDT chambers due to their position on the ATLAS toroid magnet coil. The RPC is inserted in the support frame and the sMDT chamber is mounted on kinematical bearings. The mechanical adjustment accuracy is less than 1 mm, fulfilling the requirements on the chamber alignment in the detector \cite{b1}. 

\begin{figure}[htbp]
\centerline{\includegraphics[width=8.5cm]{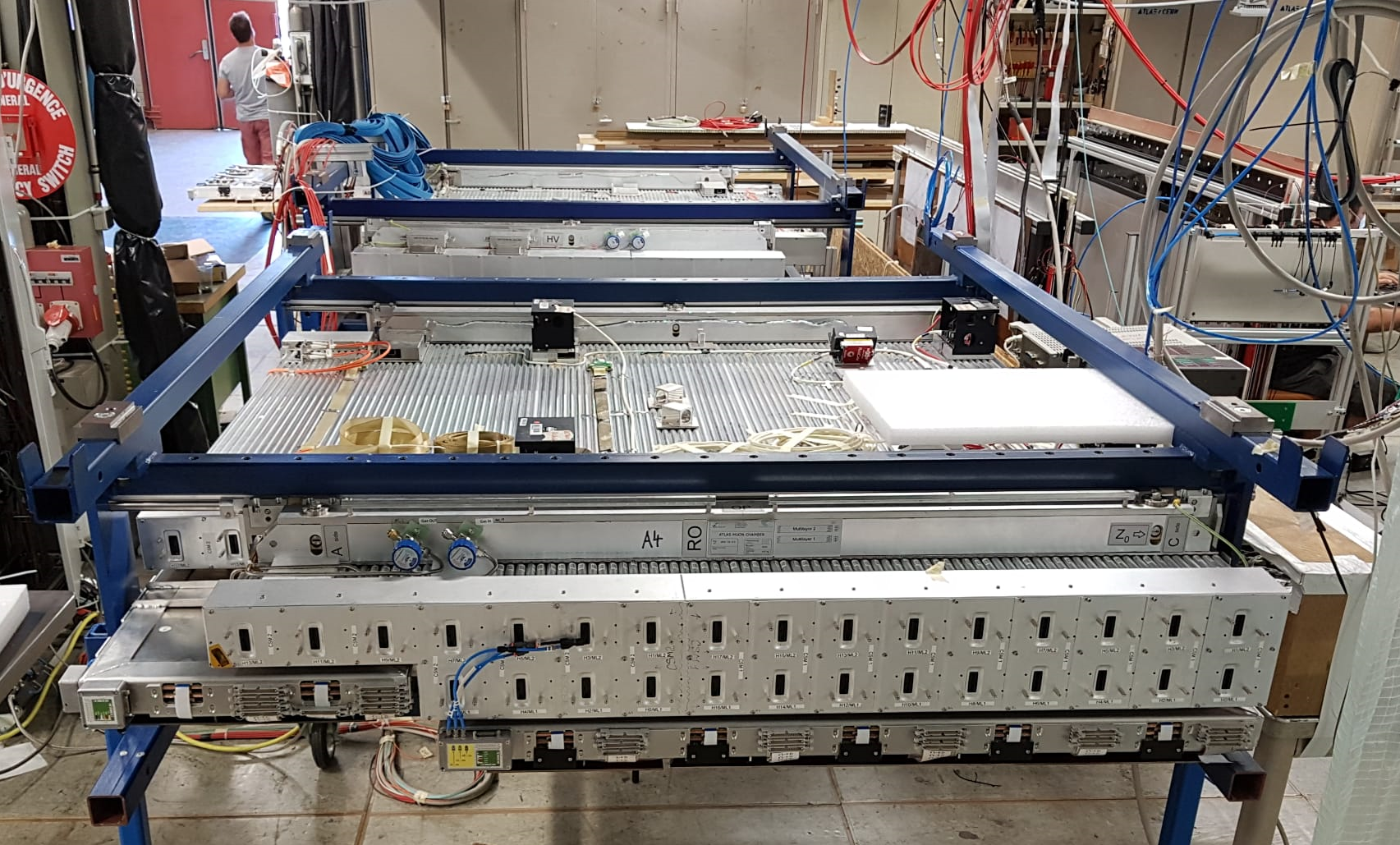}}
\caption{Final BIS78 muon station with installed B-field and Axial-Praxial, CCC alignment sensors.}
\label{fig5}
\end{figure}

In order to study the effect of the integrated RPC detectors and its electronics on the sMDT system, the interference test is performed. Then the sMDT detector is equipped with B-field and Axial-Praxial, CCC alignment sensors whose proper operation is verified. The photographs of the detectors integration process and of the final muon station are shown in Fig.~\ref{fig4} and~\ref{fig5}.

\section{Installation  and  Commissioning of new sMDT stations in the ATLAS experiment}
\begin{figure}[htbp]
\centerline{\includegraphics[width=8.5cm]{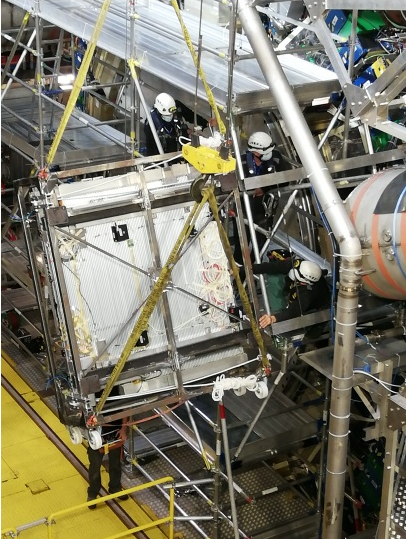}}
\caption{Photograph of the installation process of the new BIS78 muon station in the ATLAS muon spectrometer.}
\label{fig6}
\end{figure}

At the surface above the ATLAS experimental cavern, the mechanical integrity of the station is checked to ensure that no damages occurred during transport from the validation and integration site. The new muon station is placed into an installation frame and lowered into the underground experimental hall. There, the installation frame is supported by two cranes, rotated to the appropriate angle, and docked to the rail system of the ATLAS Muon spectrometer. The muon station is slid onto the rails by using the two winches, positioned with an accuracy of about a few mm in all three spatial directions and fixed at one bearing on the rail. A photograph of the BIS78 muon station installation is shown in Fig.~\ref{fig6}. Installation of the new BIS78 muon stations for the A-side of the ATLAS muon spectrometer began in mid-September 2020 and lasted until the end of January 2021. 

The underground commissioning phases are required to bring the new sMDT sub-system into an operational state suitable for future ATLAS physics data-taking campaigns. Any potential damages and performance losses are immediately identified and repaired after installation. The correct sag adjustment is verified with the on-chamber alignment system. ATLAS global alignment systems are checked as well. No failures have been observed after installation. Moreover, the detector's basic parameters, such as no leaks in the gas, high-voltage stability of the chamber with the operating gas mixture, low background noise, appropriate connectivity of the electronic components, and the operational characteristics of the front-end/back-end electronics, are set to optimal settings.

The sMDT sub-detectors have been fully added and integrated into the ATLAS operational Data Acquisition/Detector Control/Data Quality Monitoring systems. 

\section{Conclusion}
In the coming years, the  ATLAS muon spectrometer will go through a series of modernizations in order to cope with the foreseen increase in LHC performance. New BIS78 detector stations (the triplet thin gap Resistive Plate Chamber and small (15 mm) diameter Muon Drift Tube detector) have been constructed for the Phase-1 upgrade as a pilot project for the innermost barrel region of the ATLAS muon system to increase the trigger acceptance and to prepare for the high background rate requirements of the HL-LHC. The Max Planck Institute for Physics in Munich has built 16 new sMDT chambers. All sMDT detectors passed the acceptance tests at CERN and were integrated with their RPC trigger chambers. A total of 8 out of 16 muon stations were successfully installed in the ATLAS experiment before the start of the Run-3, while the muon system modernization campaign with another BIS78 modules is continued during the next third Long Shutdown (LS3) of the LHC.




\end{document}